\documentclass{pasj01}
\usepackage[switch,mathlines]{lineno}

\Received{$\langle$reception date$\rangle$}
\Accepted{$\langle$acception date$\rangle$}
\Published{$\langle$publication date$\rangle$}
\begin{document}

\title{Major merger fraction along the massive galaxy quenching channel at $0.2<z<0.7$}
\author{
Shin \textsc{Inoue}\altaffilmark{1}$^{*}$, 
Kouji \textsc{Ohta}\altaffilmark{1}, 
Yoshihisa \textsc{Asada}\altaffilmark{1,2}, 
Marcin \textsc{Sawicki}\altaffilmark{1,2}**,
Guillaume \textsc{Desprez}\altaffilmark{2},
Stephen \textsc{Gwyn}\altaffilmark{3},
Vincent \textsc{Picouet}\altaffilmark{4}
}
\altaffiltext{1}{Department of Astronomy, Kyoto University, Kitashirakawa-Oiwake-Cho, Sakyo-ku, Kyoto, Kyoto 606-8502, Japan}
\altaffiltext{2}{Department of Astronomy and Physics and Institute for Computational Astrophysics, Saint Mary's University, 923 Robie Street, Halifax NS, B3H 3C3, Canada}
\altaffiltext{3}{NRC-Herzberg, 5071 West Saanich Road,Victoria,British Columbia V9E 2E7,Canada}
\altaffiltext{4}{Cahill Center for Astrophysics, California Institute of Technology, 1216 East California Boulevard, Mail Code  278-17, Pasadena, CA 91125, USA}

\email{sinoue@kusastro.kyoto-u.ac.jp}
\altaffiltext{**}{JSPS Invitational Visiting Fellow and Canada Research Chair}

\KeyWords{galaxies: evolution --- galaxies: interactions --- galaxies: statistics}

\maketitle

\begin{abstract}
We study the major merger fraction along the massive galaxy quenching channel (traced with rest-frame $\mathrm{NUV}-r$ color) at $z=$ 0.2--0.7, aiming to examine the Cosmic Web Detachment (CWD) scenario of galaxy quenching.
In this scenario, the major merger fraction is expected to be high in green valley galaxies as compared with those in star-forming and quiescent galaxies of similar stellar mass.
We used photometry in the E-COSMOS field to select 1491 (2334) massive ($M_\ast>10^{9.5}$\,$M_\odot$) galaxies with $m_i<22$\>mag ($m_z<22$\>mag) at $z=$ 0.2--0.4 ($z=$ 0.4--0.7) in the rest-frame color range of $0.8<r-K_s<1.3$.
We define a major galaxy-galaxy merger as a galaxy pair of comparable angular size and luminosity with tidal tails or bridges, and we identified such major mergers through visual inspection of Subaru-HSC-SSP PDR 2 $i$- and $z$-band images.
We classify 92 (123) galaxies as major merger galaxies at $z=$ 0.2--0.4 ($z=$ 0.4--0.7).
The resulting major merger fraction is 5\%--6\% and this fraction does not change with galaxy color along the massive galaxy quenching channel.
The result is not consistent with the expectation based of CWD scenario as the dominant mechanism of massive galaxy quenching.  
However, there are some caveats such as 
(i) the mergers that cause quenching may lose their visible merger signatures rapidly before they enter the Green Valley,
(ii) our method may not trace the cosmic web sufficiently well, and 
(iii) because of our mass limit, most of the galaxies in our sample may have already experienced CWD events at higher redshifts than those studied here. 
Further studies with deeper data are desirable in the future.
\end{abstract}
%\pagewiselinenumbers

%section 1
\section{Introduction}
Massive galaxies are mostly classified into two populations: blue star-forming (SF) galaxies and red quiescent (i.e., without star forming activity; Q) galaxies (e.g., \cite{Arnouts+13}; \cite{Moutard+16}; \cite{Fang+18}).
It is now well accepted that galaxies can move from the SF to the Q population by shutting down their star formation, a process referred to as quenching.
While the underlying mechanisms that cause quenching are not yet fully understood, they come in two flavors \citep{Peng+10}:
(1) mass quenching, whose probability appears to increase exponentially with stellar mass, $P$(quench) $\sim e^{M_\ast}$, and which therefore disproportionately affects massive galaxies ($M_\ast>10^{9.5}$\,$M_{\odot}$);
and (2) environmental quenching, whose probability appears to be independent of mass while also operating primarily in overdense regions. 

While the so-called environmental quenching is understood to be caused by environmental mechanisms, but the cause or causes of mass quenching remain unclear.
Several mechanisms have been proposed for the quenching of massive galaxies (see  \cite{Man+18} for a review), such as AGN feedback (e.g., \cite{Best+05, Croton+06, Hopkins+06, Menci+06}), virial shock heating (e.g., \cite{Dekel+06}),  and supernova feedback (e.g., \cite{Ciotti+91}).
However, the most essential process is not yet clear.
Recently, \citet{AragonCalvo+19} proposed a new quenching scenario: Cosmic Web Detachment (CWD).
Generally, gas is supplied into a galaxy's halo by accretion along a cosmic web structure (e.g., \cite{Dekel+09}), and the galaxy maintains star formation using the accreting gas.
In the CWD scenario, initially a galaxy is connected to a web of filaments along which the gas is accreting into the halo.
Subsequently, an interaction with large scale structures or interaction between the galaxy's (dark matter) halo and another halo of comparable or larger mass detaches the galaxy's halo from the gas feeding filaments.
The detachment of the galaxy from its gas supply leads to quenching.
\citet{AragonCalvo+19} claimed that the CWD event that leads to the disconnection of the gas accreting filaments is a fundamental process and processes induced by internal activity, such as AGN feedback, could be the result of the CWD event, with their  impact only compounding the quenching effect of the more fundamental CWD event.

If CWD is caused by the interaction between galaxy halos of comparable size (major mergers), disconnection of the gas feeding filaments occurs at the moment when the halos collide with each other.
At this stage of halo merging, galaxies in the halos are still separated, and the galaxy merger begins some time after the collision of the halos.
Therefore, if we assume that star formation begins to stop soon after the CWD event, i.e., quenching starts soon after the halo interaction, then we can expect that when the galaxies begin to merge, quenching is already in progress and the galaxies at this point are simply consuming their remaining gas.

Some observational results support the CWD scenario.
Using data of the ALFALFA H\,\emissiontype{I} survey, \citet{Odekon+18} studied the relationship between H\,\emissiontype{I} deficiency in galaxies and the environment where they reside.
They found that the H\,\emissiontype{I} deficiency of galaxies with the stellar mass range of $10^{8.5\mathchar`-\mathchar`-10.5}$\,$M_\odot$ decreases with distance from large-scale filament spine, and also found that galaxies in ``tendrils'', within voids but in relatively denser environments, are gas-rich as compared to those in voids. This fits a picture that galaxies are initially connected to gas-supplying filaments and interactions with large scale structures detach the gas-feeding filaments.
In a different study, \citet{Moutard+20} investigated the distribution of X-ray AGNs along the mass-related quenching channel to unveil quenching mechanism of massive ($M_\ast>10^{10.6}$\,$M_\odot$) galaxies.
They found that heavily obscured X-ray AGNs are mostly hosted by massive galaxies in the green valley (GV).
The median X-ray hardness ratio in such galaxies corresponds to a hydrogen column density of $\sim 3\times 10^{22}$\,cm$^{-2}$, which is a typical value predicted by simulations \citep{Blecha+18} in the {\it final stage}  of {\it gas-poor} massive galaxy major mergers, i.e., galaxy major mergers that occur after the galaxies have left the star-forming main sequence. Thus, \citet{Moutard+20} suggested that a galaxy major merger is not the direct cause for the quenching but, rather, the aftermath of the mechanism that causes the quenching.
This view fits the CWD scenario framework that the detachment of the gas-feeding filament by the dark matter halo merger, which precedes the galaxy merger, is the cause of the quenching.

To test the CWD scenario and the suggestion by \citet{Moutard+20}, we will investigate the major merger fraction as function of galaxy color which corresponds to the mass-related quenching channel.
In high-resolution simulations of massive galaxy major merges (e.g., \cite{Marinacci+18, Naiman+18, Nelson+18, Pillepich+18, Springel+18, Moreno+19}), the first pericenter passage of the galaxies occurs $\sim$ 1\>Gyr after the collision of the halos.
Tidal structures induced by the galaxy interaction begin to appear at the pericenter passage and are expected to remain present  for a period of $\sim$ 1--2\>Gyr.
Color evolution of an $e$-folding star formation history ($\tau=$ 0.5--2\>Gyr) shows that a galaxy is in the SF main sequence during the first $\sim$ 1\>Gyr just after the star formation stops and is expected to spend 1--3.5\>Gyr crossing the GV region \citep{Moutard+16}.
Combining these estimates, if a star-forming activity stops at the moment of the CWD event, tidal structures are expected to be visible when the galaxy enters the GV.
That is, the fraction of galaxy-galaxy major mergers is expected to be high in the GV region in the CWD scenario.

In this paper, we intend to test Cosmic Web Detachment scenario by examining the fraction of galaxy major mergers along the mass-related quenching channel.
Using deep and high-resolution Subaru-HSC images, we identify major mergers using their morphological signatures, and examine how the major merger fraction changes along the mass-related quenching channel.
We use a flat cosmological parameter set; $\Omega_{\mathrm{m}} = 0.3$, $\Omega_{\mathrm{\Lambda}}=0.7$, and $H_0=70$\>km\>s$^{-1}$\>Mpc$^{-1}$.
The AB magnitude system is used throughout.

%section 2
\section{Data}
%section2-sub1
\subsection{Photometric data and physical parameters}
We used a galaxy catalog in the E-COSMOS field based on the combined CLAUDS and HSC-SSP surveys \citep{Desprez+23}.
This field is covered by CLAUDS ($u$ and $u^\ast$ bands; \cite{Sawicki+19}) and HSC-SSP ($g$, $r$, $i$, $z$, $y$ bands; \cite{Aihara+19}), and is partly (1.5\>deg$^2$ in the central part of the E-COSMOS field) covered by the Ultra VISTA deep NIR survey ($Y$, $J$, $H$, $K_s$ bands; \cite{McCracken+12}).
In this study, we used the catalogued galaxies in the 1.5\>deg$^2$ area covered by Ultra VISTA.
\citet{Desprez+23} present two versions of the catalog obtained with different photometry and photometric redshift codes:
One version uses HSC pipeline hscPipe and Phosphoros to conduct photometry and template-fitting, respectively.
The other version uses SExtractor and Le Phare, respectively. In this study, we employ the SExtractor plus LePhare version.
Physical parameters were measured for this catalog in \citet{Picouet+23} and we used their photometric redshifts, stellar masses and rest-frame absolute magnitudes in our work.

Advantages of the catalog are its depth of the photometry and accuracy of derived photometric redshifts.
Median $5\sigma$ depth within a \timeform{2"} diameter aperture of CLAUDS $u$ and $u^\ast$ band is $m_{u,u^\ast}=27.1$\>mag, and $m_{u,u^\ast}=27.7$\>mag in deeper area in a part of E-COSMOS field.
Median $5\sigma$ depth with a \timeform{2"} diameter aperture of Deep and Ultra Deep field of Subaru HSC-SSP Public data release 2 (PDR2) in $g$, $r$, $i$, $z$, and $y$ bands are $m_g = 27.0$\>mag, $m_r = 26.6$\>mag, $m_i = 26.4$\>mag, $m_z = 26.0$\>mag and $m_y = 25.0$\>mag.
The $5\sigma$ depth with a \timeform{2"} diameter aperture of the Ultra VISTA survey are $m_Y=24.6$\>mag, $m_J=24.4$\>mag, $m_H=23.9$\>mag, and $m_{K_s}=23.7$\>mag.
Thanks to the depth of the data and the presence of $u$, $u^\ast$ bands, the photo-$z$ accuracy is very good, with $\sigma_z/(1+z)\sim0.027$ (see \cite{Desprez+23}).

%section2-sub2
\subsection{HSC-SSP image}
We used Subaru-HSC-SSP PDR2 images \citep{Aihara+19} to conduct morphological inspections.
We selected galaxies in the redshift range $z=$ 0.2--0.7 and performed our analysis within two sub-ranges of $z=$ 0.2--0.4 and $z=$ 0.4--0.7.
We used $i$- and $z$-band images, for redshifts $z=$ 0.2--0.4 and $z=$ 0.4--0.7, respectively, which allowed us to see galaxies in the approximately rest-frame $r$-band.
The images are very deep (see above) and have excellent seeing size (\timeform{0".62} in $i$-band images and \timeform{0".63} in $z$-band images) with a pixel scale of \timeform{0".168}, enabling us to detect very low surface brightness and spatially fine tidal structures of merging galaxies.

%section 3
\section{Sample selection}
\begin{figure}
 \begin{center}
  \includegraphics[width=80mm]{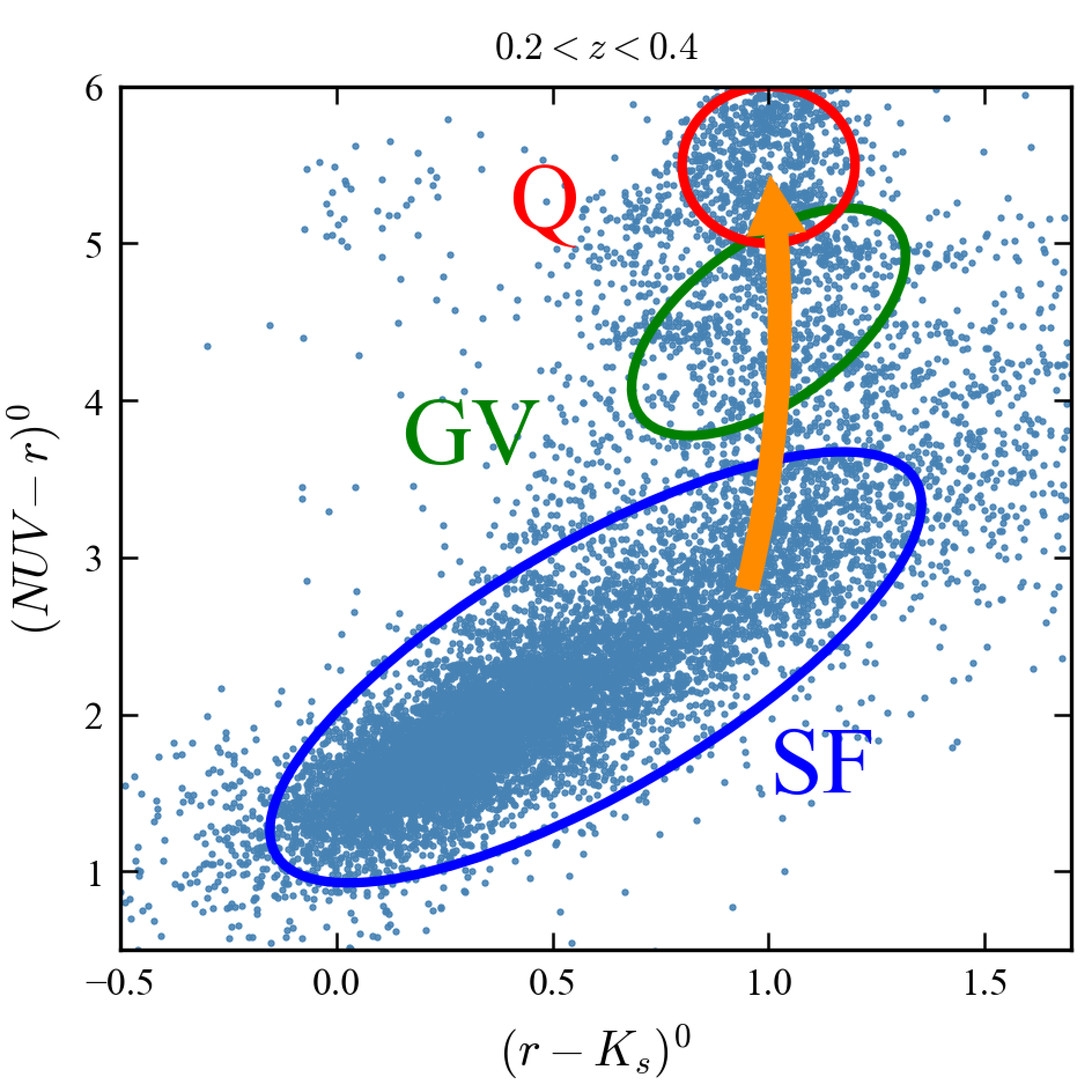}
 \end{center}
 \caption{
The NUVrK diagram.
Horizontal axis and vertical axis show rest-frame $(r-K_s)^0$ and rest-frame $(\mathrm{NUV}-r)^0$, respectively.
Galaxies at $z=$ 0.2--0.4 with $m_i<23.5$\>mag within a $3^{\prime\prime}$ diameter aperture are plotted with blue dots.
Star-forming (SF), green-valley (GV) and quiescent (Q) galaxies are indicated with blue, green and red ellipses, respectively.
The orange arrow shows the path of the mass-related quenching channel (see text).
}
 \label{NUVrK}
\end{figure}

The two-color diagram of rest-frame $(\mathrm{NUV}-r)^0$ vs $(r-K_s)^0$ (NUVrK diagram; \cite{Arnouts+13}) is a powerful tool to separate the galaxy population into star-forming, quiescent, and green valley galaxies.
The NUVrK diagram is similar to the well-known UVJ diagram (\cite{Williams+09}), but its longer y-axis wavelength baseline allows for a clearer identification of the transitional green valley region.
In the NUVrK diagram, SF galaxies are distributed in a region with $0 \lesssim (r-K_s)^0 \lesssim 1.5$ and $1 \lesssim (\mathrm{NUV}-r)^0 \lesssim 3.5$, making a sequence in color-color space (figure \ref{NUVrK}).
The quiescent galaxy population is distributed in a region with $ 0.7 \lesssim (r-K_s)^0 \lesssim1.3$ and  $ 5 \lesssim (\mathrm{NUV}-r)^0 \lesssim 6$  (figure \ref{NUVrK}).

Galaxies that reside in the region between the SF and Q populations are GV galaxies (figure \ref{NUVrK}).
Since GV galaxies show specific star formation rates between SF galaxies and quiescent galaxies \citep{Moutard+20}, GV galaxies are expected to be related to the quenching phase of evolution.
\citet{Moutard+16} found that most of the GV region is populated by massive galaxies (60 percent of the GV galaxies have stellar mass of $10^{10.5\mathchar`-\mathchar`-11}$\,$M_\odot$), connecting them to the idea of mass quenching \citep{Ilbert+10, Peng+10}.
They also found that distribution of GV galaxies is concentrated in the narrow color slice of $0.76\lesssim(r-K_s)^0\lesssim1.23$, and concluded that if the stellar mass of SF galaxies exceeds characteristic value ($M_\ast\sim10^{10.64}$\,$M_\odot$), they start to stop their star formation activities and move upward to through relatively narrow color space channel (called ``mass related quenching channel''; orange arrow in figure \ref{NUVrK}) in the NUVrK diagram, from SF galaxies through GV galaxies to Q galaxies.
Galaxies quenching through the more rapid environmental quenching processes also move from the SF region to the Q region via the GV; however, they move through the GV following a track that's bluer in $(r-K_s)^0$ than that followed by the mass-quenching galaxies, and their rapid GV transition timescales mean that this region of the GV is less populated with transitioning galaxies given the resulting lower visibility timescales \citep{Moutard+16,Moutard+18}.

%figure1

\begin{figure*}
 \begin{center}
  \includegraphics[width=160mm]{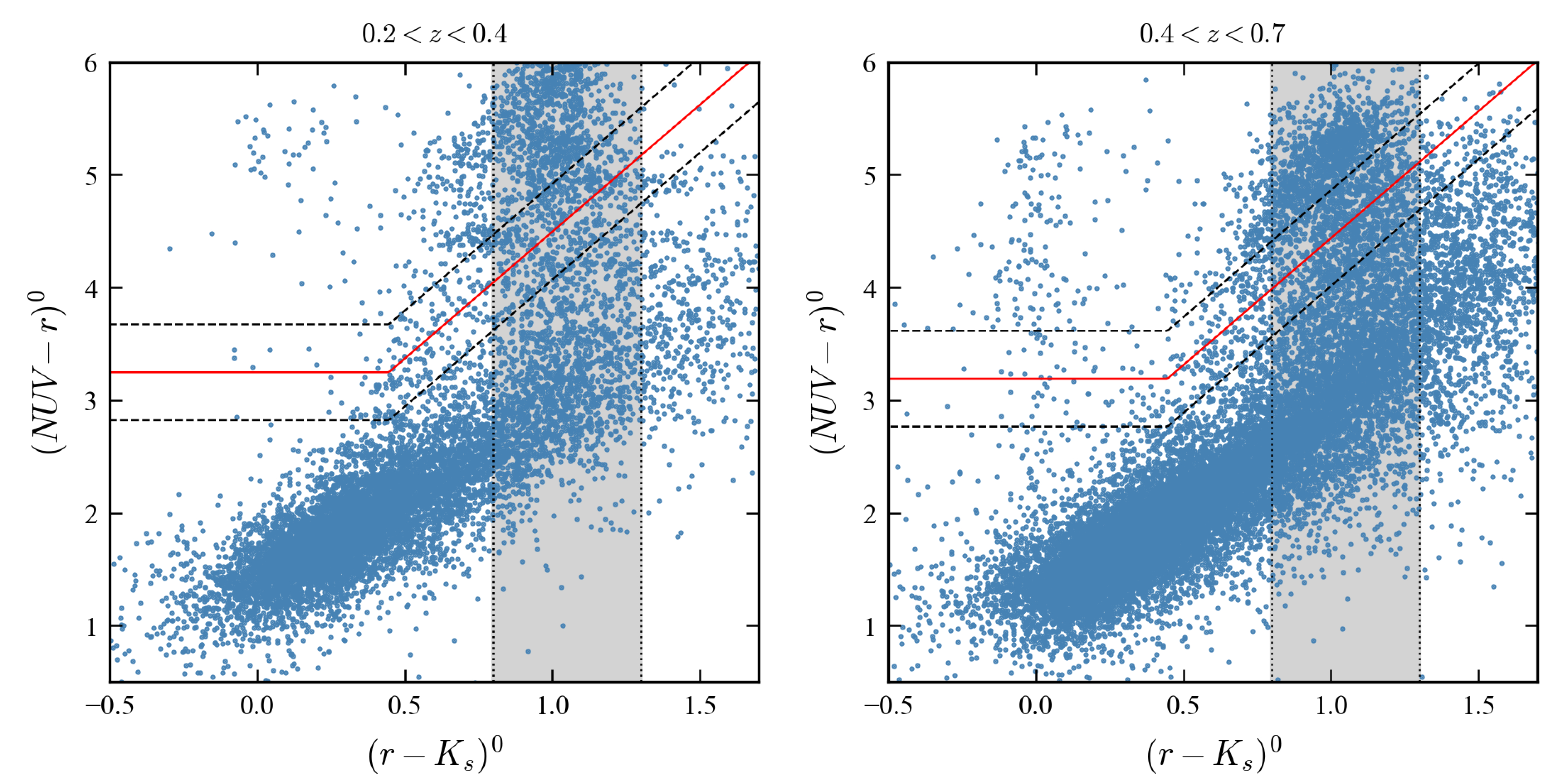}
 \end{center}
 \caption{
NUVrK diagram.
The horizontal axis and the vertical axis show rest-frame $(r-K_s)^0$ and rest-frame $(\mathrm{NUV}-r)^0$, respectively.
Galaxies at $z=$ 0.2--0.4 with $m_i<23.5$\>mag within a \timeform{3"} diameter aperture (left) and galaxies at $z=$ 0.4--0.7 with $m_z<23.5$\>mag within a $3^{\prime\prime}$ diameter aperture (right) are plotted with blue dots.
We selected galaxies in the gray shaded zone between the two dotted vertical lines ($0.8<(r-K_s)^0 <1.3$).
Black dashed lines show upper/lower limit of GV region defined by \citet{Moutard+20} and the galaxies in the region between them are GV galaxies. 
Red solid lines refer to the central value of $\mathrm{NUV}-r$ color of the GV region defined by equation (\ref{GV_central}).
}
\label{NUVrK_track}
\end{figure*}

In figure \ref{NUVrK_track}, we plot all the cataloged galaxies at $z=$ 0.2--0.4 with $m_i<23.5$\>mag within a $3^{\prime\prime}$ diameter aperture (left), and at $z=$ 0.4--0.7 with $m_z<23.5$\>mag within a $3^{\prime\prime}$ diameter aperture (right).
The sequence of blue SF galaxies and the clump of red quiescent galaxies are clearly seen.
The region of GV defined by \citet{Moutard+20} is shown in figure \ref{NUVrK_track} as the region between two dashed lines.
We defined $(r-K_s)^0$ color range of $0.8<(r-K_s)^0 <1.3$ as the massive galaxy quenching channel to fit our data (gray band between two dotted lines in figure \ref{NUVrK_track}).
In this study, we focus on the massive galaxy quenching channel and select the galaxies in the color range of $0.8<(r-K_s)^0 <1.3$.

%figure2

\begin{figure*}
 \begin{center}
  \includegraphics[width=160mm]{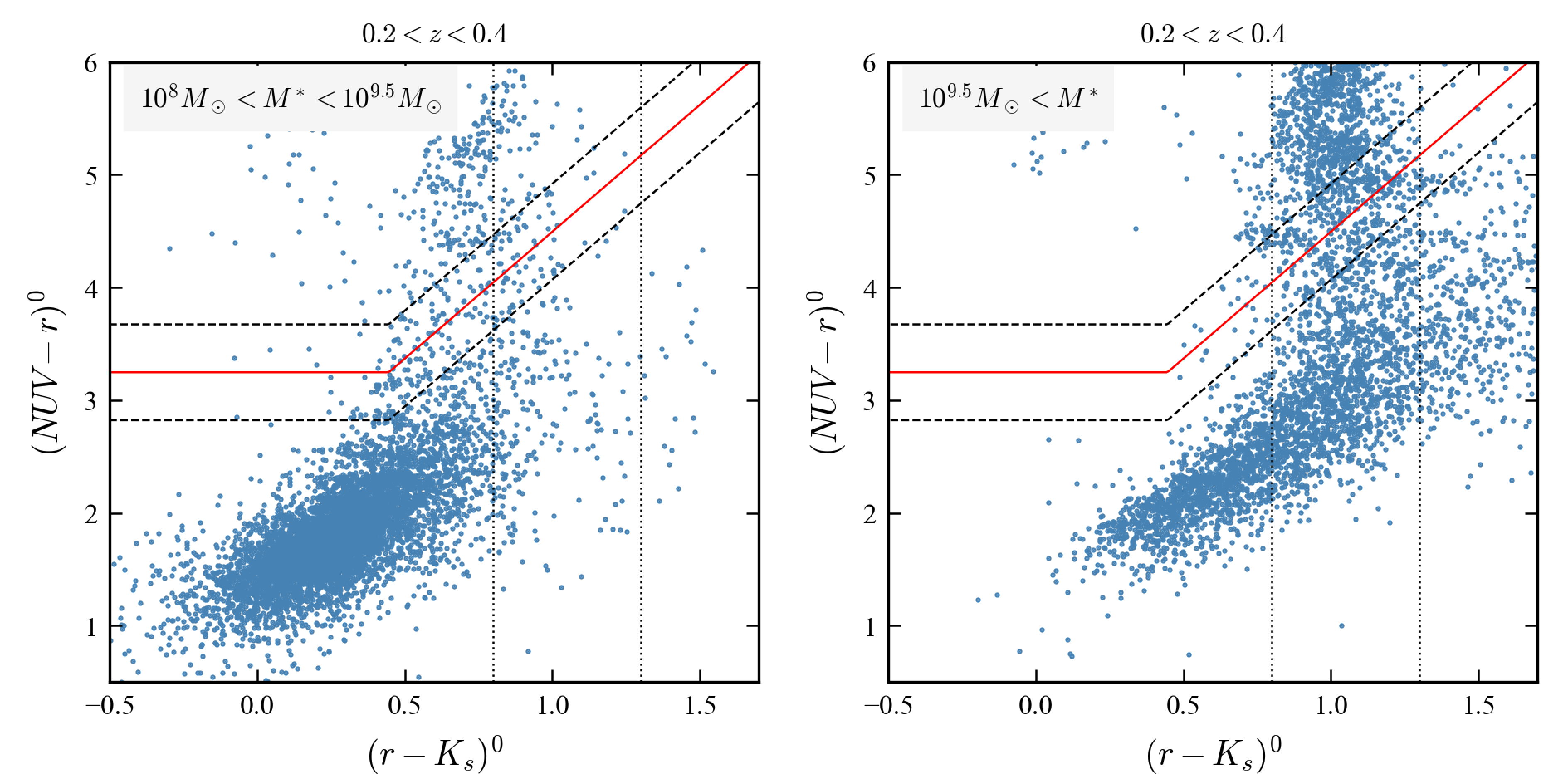}
 \end{center}
  \caption{
Same as the left panel of figure \ref{NUVrK_track}, but for galaxies with $10^8$\,$M_\odot < M_\ast < 10^{9.5}$\,$M_\odot$ (left panel) and  $10^{9.5}$\,$M_\odot < M_\ast$ (right panel).
}
  \label{NUVrK_mass}
\end{figure*}

\begin{figure*}
 \begin{center}
  \includegraphics[width=160mm]{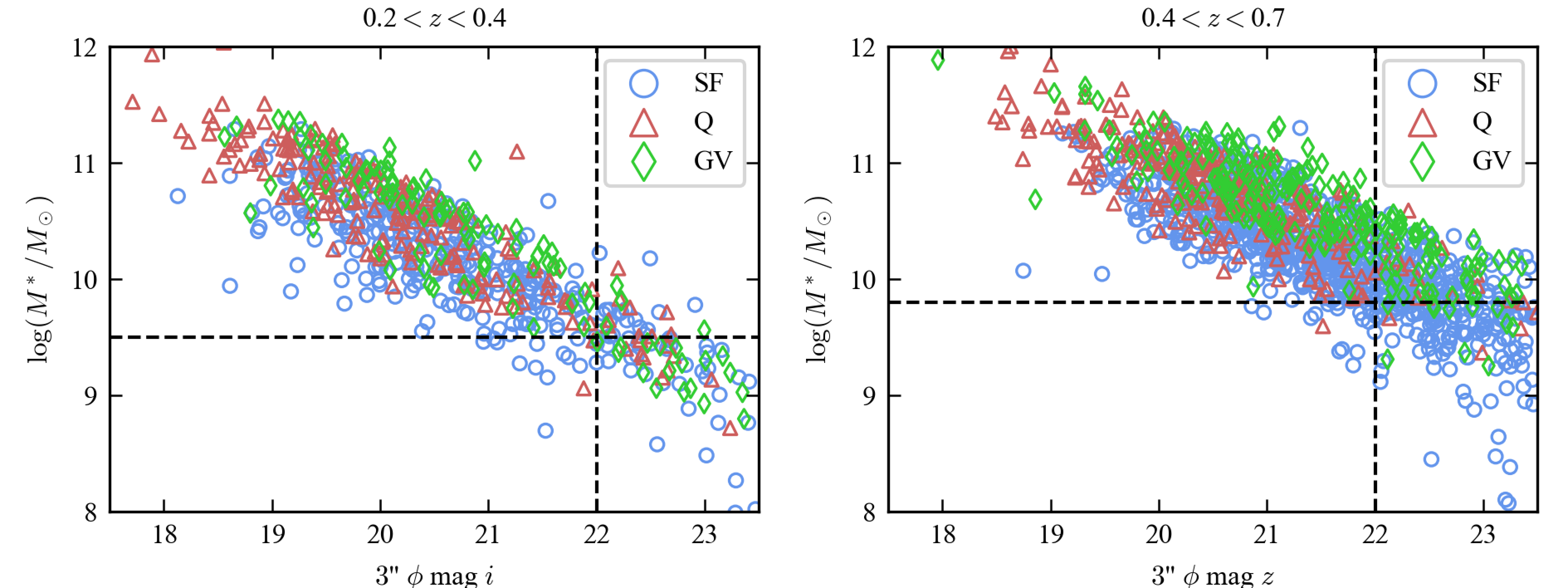}
 \end{center}
  \caption{
Stellar mass versus apparent magnitude.
Galaxies in the color range of $0.8<(r-K_s)^0<1.3$ are plotted.
Left panel is for the redshift range of $z=$0.2--0.4; the horizontal axis shows $i$-band magnitude within a \timeform{3"} aperture.
Right panel shows for the redshift range of $z=$0.4--0.7; the horizontal axis shows $z$-band magnitude within a \timeform{3"} aperture.
The black dashed lines show the stellar mass and apparent magnitude limits (see text).
Blue circles, green diamonds, and red triangles denote the SF, GV and Q galaxies, respectively (see figure \ref{NUVrK_track}).
}
  \label{appmag_mass}
\end{figure*}

Figure \ref{NUVrK_mass} shows the NUVrK diagrams for the galaxies in the left panel of figure \ref{NUVrK_track} ($z=$ 0.2--0.4) but now divided into two stellar mass ranges.
The left panel shows the distribution of galaxies with a stellar mass of $10^8$\,$M_\odot<M_\ast<10^{9.5}$\,$M_\odot$ and the right panel shows that of the  galaxies with a stellar mass of $M_\ast > 10^{9.5}$\,$M_\odot$.
Most of the galaxies with $M_\ast < 10^{9.5}$\,$M_\odot$ are located outside of the color range of $0.8<(r-K_s)^0<1.3$, consistent with the idea of rapid, environmental quenching \citep{Moutard+18} that affects primarily lower-mass galaxies \citep{Peng+10}.
In contrast, galaxies with $M_\ast>10^{9.5}$\,$M_\odot$ are in the massive galaxy quenching channel ($0.8<(r-K_s)^0 <1.3$), as expected.
In this study we focus on massive galaxy quenching channel, and therefore we selected the galaxies with $M_\ast>10^{9.5}$\,$M_\odot$ for the redshift range of $z=$0.2--0.4.
In addition, since it is difficult to detect tidal features if their parent galaxies are faint, we also imposed an apparent magnitude limit selection criterion set to $m_i<22$\>mag within a \timeform{3"} diameter aperture.
The distribution of apparent magnitude against stellar mass at $z=$ 0.2--0.4 is shown in the left panel of figure \ref{appmag_mass}.
As seen in figure \ref{appmag_mass} (left panel), the apparent magnitude limit of $m_i<22$\>mag excludes the low-mass GV galaxies of $M_\ast<10^{9.5}$\,$M_\odot$ at $z=$ 0.2--0.4, which is fine as our goal is to study the quenching of massive galaxies.
In the $z=$ 0.4--0.7 redshift range, we imposed similar but slightly modified criteria: stellar mass with $M_\ast > 10^{9.8}$\,$M_\odot$ and apparent magnitude with $m_z<22$\>mag (right panel of figure \ref{appmag_mass}).

%figure3
%figure4

\citet{Moutard+20} provide definitions of the extent of the GV, which we adopt and reproduce with dashed lines in figures \ref{NUVrK_track} and \ref{NUVrK_mass}.
Using equation (1) in \citet{Moutard+20}, we calculate the central value of $(\mathrm{NUV}-r)^0$ of the GV extent as function of color $(r-K_s)^0$ as
%equation1
\begin{equation}
\label{GV_central}
(\mathrm{NUV}-r)^0_{\mathrm{center}}\\
=
\left\{
\begin{array}{l}
3.347 -0.029 t_{\mathrm{L}}\\
\quad\quad\quad\quad \mathrm{if~} (r-K_s)^0<0.4462,\\
2.25 (r-K_s)^0+2.343-0.029 t_{\mathrm{L}}\\
\quad\quad\quad\quad \mathrm{if~}(r-K_s)^0>0.4462,
\end{array}
\right.
\end{equation}
where $t_{\mathrm{L}}$ is lookback time in Gyr and the subscript ``center'' refers to the center of the GV, as marked by the red solid line.
We used the values of lookback time as $t_L=3.418$\>Gyr and $t_L=5.385$\>Gyr, which correspond to $z=0.3$ and $z=0.55$, for $z=$ 0.2--0.4 and $z=$ 0.4--0.7, respectively.

In order to track galaxy positions along the massive galaxy quenching channel, we follow \citet{Noirot+22} and define $\delta\mathrm{GV}$ as the $(\mathrm{NUV}-r)^0$ color offset from the central color of the GV.
Then, $\delta$GV is defined as the difference of $(\mathrm{NUV}-r)^0$ from $(\mathrm{NUV}-r)^0_\mathrm{center}$,
%equation2
\begin{equation}
\label{eq:delta_GV}
\delta \mathrm{GV}\equiv (\mathrm{NUV}-r)^0_{\mathrm{object}} - (\mathrm{NUV}-r)^0_{\mathrm{center}},
\end{equation}
at a galaxy's $(r-K_s)^0$ position,  $(r-K_s)^0_{\mathrm{object}}$.
I.e., $\delta\mathrm{GV}$ is the vertical offset from the red solid lines in figures \ref{NUVrK_track} and \ref{NUVrK_mass}. 

\begin{figure*}
 \begin{center}
  \includegraphics[width=160mm]{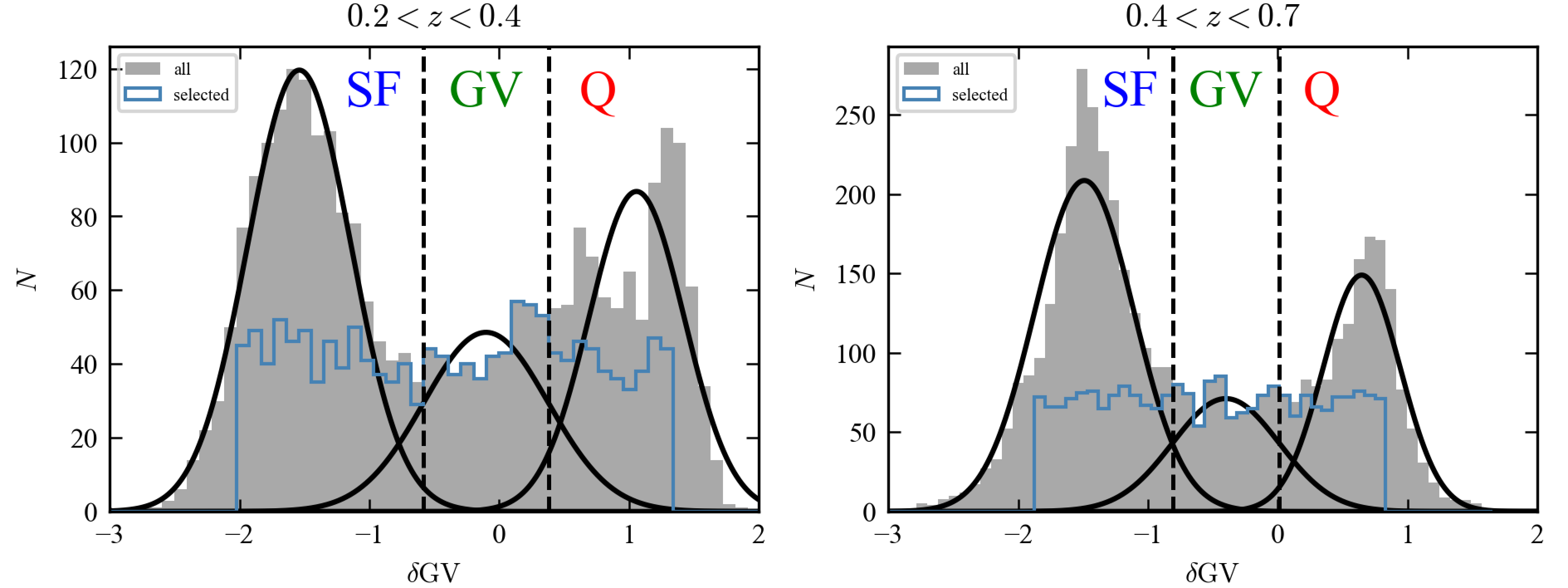}
 \end{center}
 \caption{
Histograms of $\delta$GV values for galaxies in the color range of $0.8<(r-K_s)^0 <1.3$.
The gray histograms show galaxies at $z=$ 0.2--0.4 with $m_i < 22$\>mag and $M_\ast>10^{9.5}$\,$M_\odot$ (left) and those at $z=$ 0.4--0.7 with $m_z < 22$\>mag and $M_\ast>10^{9.8}$\,$M_\odot$ (right).
The three Gaussians are the decomposed populations of SF, GV, and Q galaxies. Vertical dashed line shows the $\pm1\sigma$ of GV population.  Blue histogram shows the distributions of the galaxies that were randomly selected for visual inspection.
}
\label{d_GV_hist}
\end{figure*}

Figure \ref{d_GV_hist} shows the histogram of $\delta$GV values within the massive galaxy quenching channel at $z=$ 0.2--0.4 (left) and $z=$ 0.4--0.7 (right).
The left hump and the right hump in each panel consist of the SF galaxies and the Q galaxies, respectively. The color range between these two humps corresponds to the GV region.  Figure \ref{d_GV_hist} indicates that $\delta \mathrm{GV}$ continuously tracks color change of galaxies along the massive galaxy quenching channel. Assuming that the distribution of each of three populations can be described by a Gaussian, we decomposed the distribution (figure \ref{d_GV_hist}). We defined the GV galaxies as in the color range within $\pm 1\sigma$ centered at the mean value of $\delta$GV for the GV population; $-0.58<\delta \mathrm{GV} < 0.38$ and $-0.81<\delta\mathrm{GV}<0.01$ at $z=$ 0.2--0.4 and $z=$ 0.4--0.7, respectively, as shown with
vertical dashed lines in figure \ref{d_GV_hist}. 

%figure5

\begin{figure*}
 \begin{center}
  \includegraphics[width=160mm]{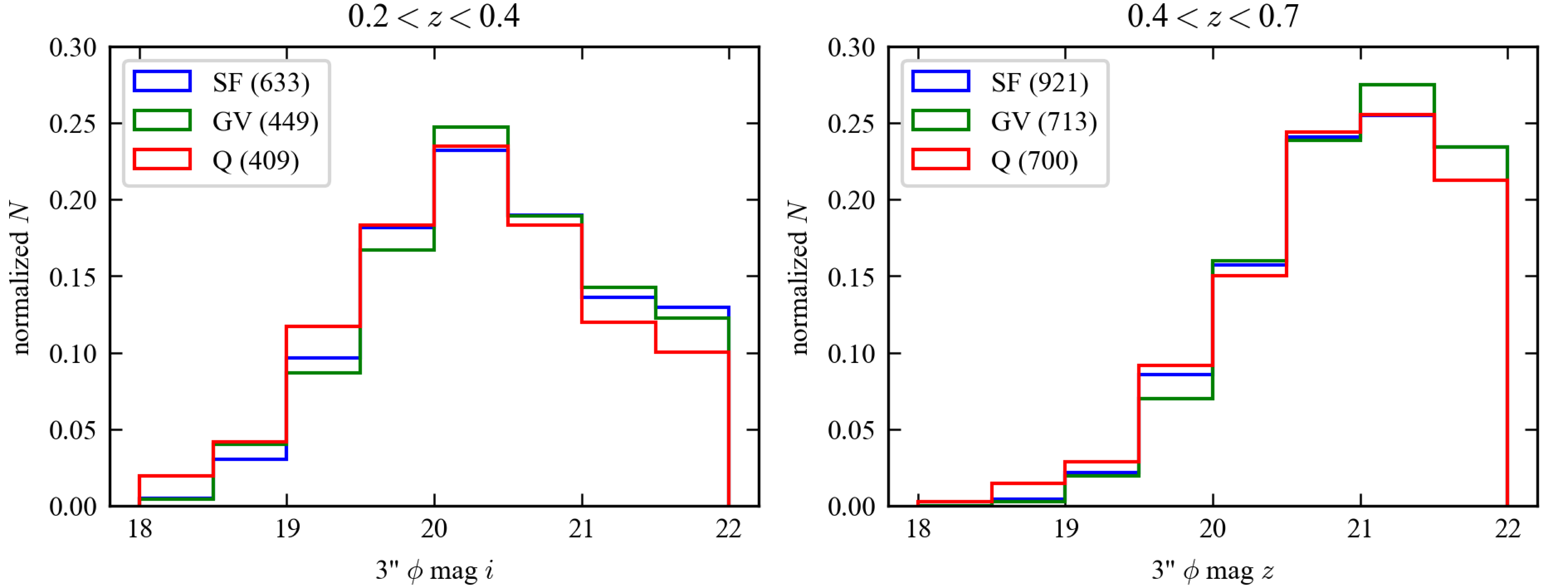}
 \end{center}
  \caption{
Normalized distribution of apparent magnitude of selected galaxies. 
The numbers of the galaxies in each plot are indicated in the legends.
Left panel shows for the redshift range of $0.2<z<0.4$; the horizontal axis shows $i$-band magnitude with a \timeform{3"} aperture.
Right panel shows for the redshift range of $0.4<z<0.7$; the horizontal axis shows $z$-band magnitude with a \timeform{3"} aperture.
SF, GV, and Q galaxies are colored with blue, green, and red, respectively.
}
  \label{mag_dist}
\end{figure*}

In the color range of GV (the region between two vertical dashed lines in figure \ref{d_GV_hist}), there are 449 and 713 galaxies at $z=$ 0.2--0.4 and $z=$ 0.4--0.7, respectively, and we inspected morphologies for all of them. For SF and Q galaxies, which are more numerous, our aim was to inspect randomly-chosen sub-samples with similar size and apparent magnitude distributions to that of the GV galaxies.
To this end, for the SF galaxies, we first extracted galaxies in the $\delta \mathrm{GV}$ range of $-2.1 < \delta \mathrm{GV} <- 0.58$ ($-1.88 < \delta \mathrm{GV} < -0.81$) for $z=$ 0.2--0.4 ($z=$ 0.4--0.7).
For the Q galaxies, we extracted galaxies in the $\delta \mathrm{GV}$ range of $0.38 < \delta \mathrm{GV} < 1.35$ ($0.01 < \delta \mathrm{GV} < 0.83$) for $z=$ 0.2--0.4 ($z=$ 0.4--0.7).
We then divided the range of $\delta$GV into smaller bins and  we divided the galaxies in each bin by their $m_i$ or $m_z$ for the redshift range of $z=$ 0.2--0.4 or $z=$ 0.4--0.7, respectively.
In each $\delta$GV bin and range of apparent magnitude, we randomly selected almost the same number of galaxies as the average number of GV galaxies in the apparent magnitude range and width of the $\delta$GV bin.
The numbers of the selected SF, GV, and Q galaxies are 633 (921), 449 (713), and 409 (700) at $z=$ 0.2--0.4 ($z=$ 0.4--0.7), respectively.
The blue histograms in figure \ref{d_GV_hist} show the distributions of the selected galaxies as function of $\delta$GV, and figure \ref{mag_dist} shows the distribution of apparent magnitude of selected samples.  As these figures show, the galaxies selected for visual inspection have a uniform distribution in $\delta$GV, and their apparent magnitude distributions are well matched.

%figure6

In our sample selection, we did not consider the effect of contamination of the color of host galaxies by AGN emission.
Here, we note that Chandra observed the COSMOS field to limiting depths of $2.2\times10^{-16}$, $1.5\times10^{-15}$, and $8.9\times10^{-16}$\>erg\>cm$^{-2}$ s$^{-1}$ in the 0.5--2\>keV, 2--10\>keV, and 0.5--10\>keV bands, respectively \citep{Civano+16}.
We matched the X-ray sources to our sample and found that only $\sim$ 3.5\%--3.6\% of the selected samples were detected in X-rays.
X-ray luminosities in the 2--10\>keV band of almost all of the matched sources are $<10^{40}$\>erg\>s$^{-1}$ at $z\sim0.4$.
Thus, the effect of AGN on galaxy colors is considered to be negligible.

%Section 4
\section{Merger inspection}
We visually inspected all of the selected galaxies to identify major mergers.
Although automated approaches such as machine learning (e.g., \cite{Goulding+18}; \cite{Thibert+21}; \cite{Omori+23}) have been used recently, visual merger identification is nevertheless highly suitable for our purposes.
One of the reasons for this is that our sample is small enough ($\sim 4000$) so that visual inspection is efficient as compared to the effort needed to set up a machine-learning tool.
Another reason is that this study is a pilot project and visual inspection allows us to learn what merger features look like in the dataset we use.
Visual inspection thus prevents us from missing some mergers that we could miss if we employed an automated classifier from the start.  Finally, our sample of visually-identified mergers can provide a training set for future machine-based classification approaches.

\begin{figure*}
 \begin{center}
  \includegraphics[width=160mm]{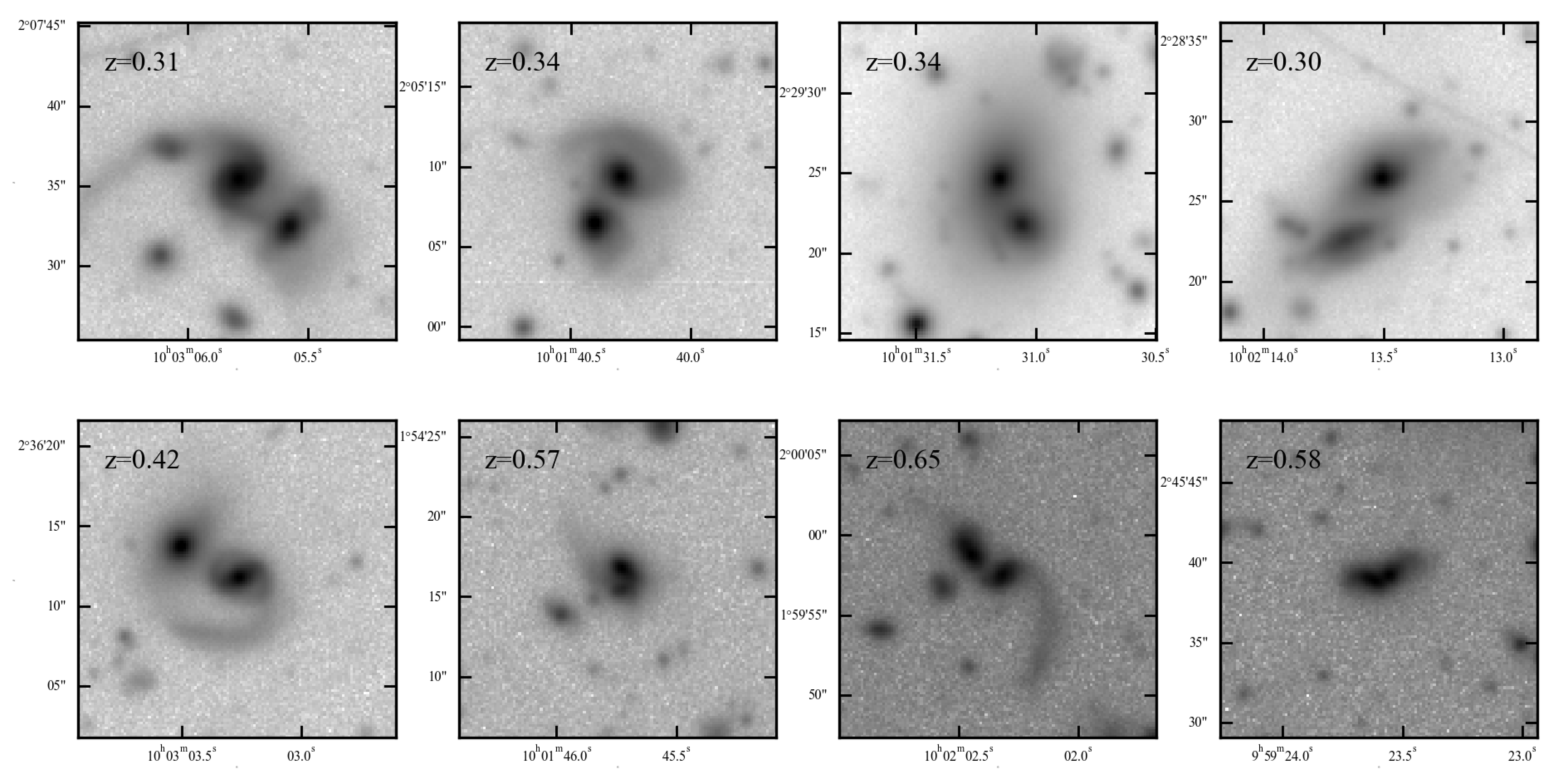}
 \end{center}
 \caption{
Examples of galaxies classified as a major merger. 
North is at the top and east to the left. A field of view of \timeform{20"}$\times$\timeform{20"} is shown.
The photometric redshift of the galaxy is shown at the upper left corner of each image.
Upper row shows an example of the galaxies in the redshift range of $0.2<z<0.4$, and Subaru HSC $i$-band images are shown.
Lower row shows an example of the galaxies in the redshift range of $0.4<z<0.7$, and Subaru HSC $z$-band images are shown.
}
\label{example_of_major_merger}
\end{figure*}

\begin{figure}
 \begin{center}
  \includegraphics[width=80mm]{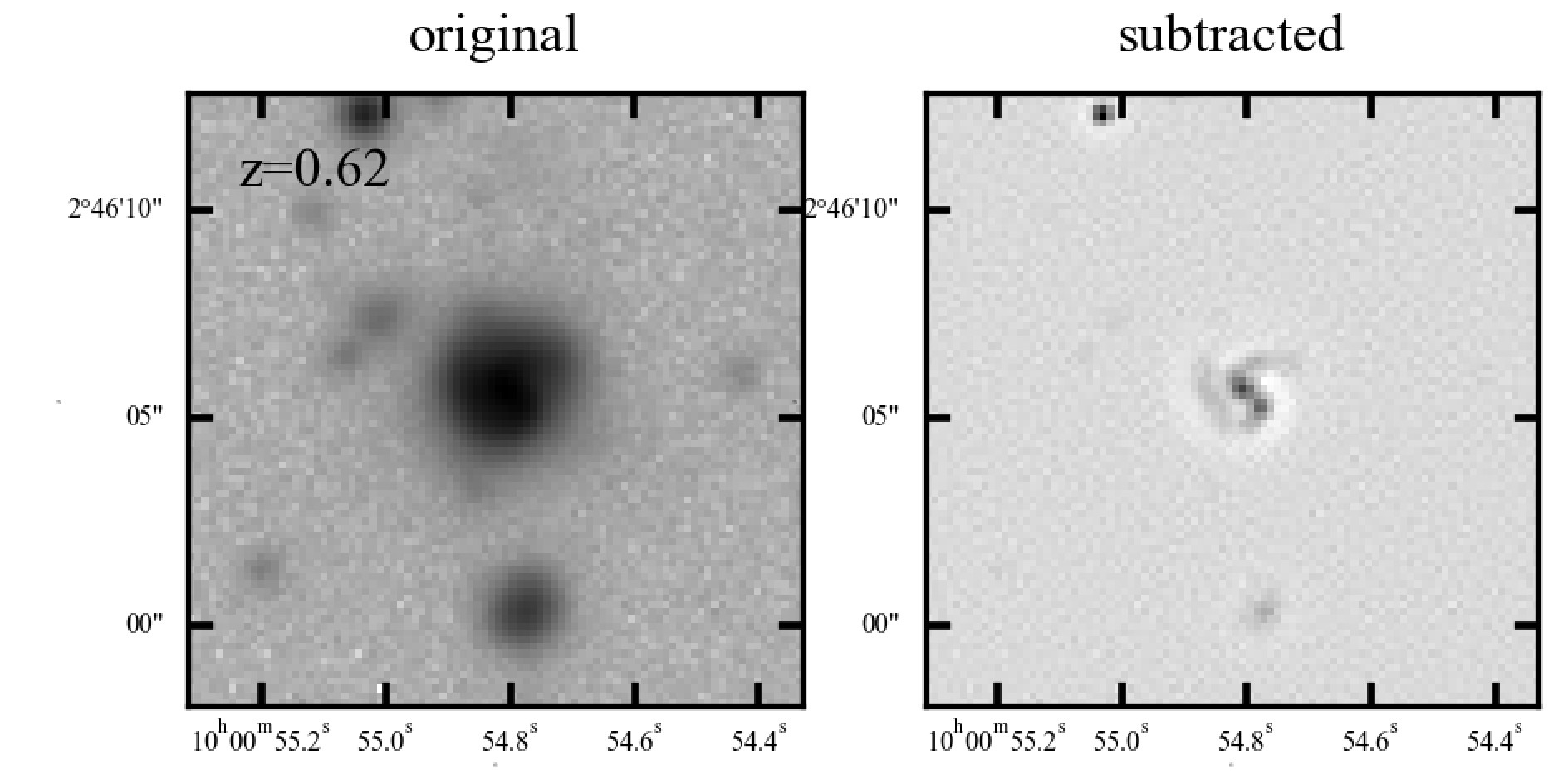}
 \end{center}
 \caption{
Example of galaxy with double nuclei.
Left panel shows the original image and right panel shows subtracted image.
North is at the top and east to the left. A field of view of \timeform{15"}$\times$\timeform{15"} is shown.
Redshift of the galaxy is $z=0.62$ and Subaru HSC $z$-band image is shown.
}
\label{smooth_double_nuclei}
\end{figure}

In our visual inspection, Subaru HSC $i$- and $z$-band images are used to identify mergers in the redshift range of $z=$ 0.2--0.4 and $z=$ 0.4--0.7, respectively.
Our criterion for classifying a galaxy as undergoing a  major merger is straightforward:
\begin{itemize}
\item A galaxy that belongs to a galaxy pair of comparable angular size and luminosity with tidal tails or bridges.
\end{itemize}
Examples of such mergers are shown in figure \ref{example_of_major_merger}.
However, this criterion alone tends to miss galaxies in the final coalescence stage, because two galaxies in this stage are expected to be close to each other. In such cases, visual inspection of direct images may identify them as a single, non-interacting galaxy.
In order to avoid such misidentifications, we supplemented our direct images with images created by subtracting a smoothed image convolved with a gaussian kernel with standard deviation of 0.5 pix from the original image.
Inspecting these processed images enables us to find a double nuclei clearly. Figure \ref{smooth_double_nuclei} shows an example of original and subtracted images and it is clear that the two close nuclei are visible in the subtracted image.

%figure7
%figure8

In our visual inspection, initial calibration to identify merger galaxies was done by 3 people (SI, YA, and MS) using $\sim500$ galaxies to achieve consistent results and to avoid possible biases when only one author inspects the sample galaxies.
After that, SI classified the selected 3825 galaxies.
A key point is that the visual inspection is ``blind'' when it comes to where the galaxy is in the quenching track -- i.e., the classifier does not know the galaxy's $\delta\mathrm{GV}$ value.
Furthermore, the classification is done on monochromatic images and thus avoids giving any clue to the classifier as to the quenching stage of the galaxy.
This approach ensures that the results are not biased by the classifier's preconceptions about where along the training track mergers should occur.

Our classification has identified 92 and 123 ($\sim$ 5\%--6\%) galaxies as major mergers at $z=$ 0.2--0.4 and $z=$ 0.4--0.7, respectively.
Among them, 11 and 22 galaxies at $z=$0.2--0.4 and $z=$ 0.4--0.7, respectively, show double nuclei identified in the smoothed-subtracted images.
Comparison with the Chandra catalog shows that only 3.7\%--4.2\% of our major merger galaxies are detected in X-rays.
The fractions of X-ray merger are almost the same in SF, GV, and Q, though the statistical significance is very low.

Although there have been many studies of the major merger fraction, it is very difficult to directly compare our value to those in previous studies, because the fraction depends on the adopted criteria for major merger, sample properties, quality of images, etc.
In this respect, the result by \citet{Bridge+10} would be suitable for comparison. These authors focused on massive ($M_\ast>10^{9.5}$\>$M_\odot$) and bright ($i_{\mathrm{vega}}<22.2$\>mag) galaxies useing CFHTLS-Deep $i'$ band images and visual inspection  to identify tidal tails, bridges, and double nuclei. 
Their reported merger fraction is $\sim$ 4\%--7\% in the redshift range of $z\sim$ 0.2--0.7.
Although the images they used are shallower than our Subaru images, their merger fractions are consistent with our values.

Another suitable comparison is with  \citet{Thibert+21}, who used Subaru HSC-SSP images (from the internal data release intermediate between PDR1 and PDR2) which are almost the same depth and resolution as HSC-SSP PDR2 images used in this paper.
These authors focused on the massive ($M_\ast>10^{10.5}$\>$M_\odot$) and bright ($r_{\mathrm{AB}}<23.0$\>mag) galaxies, and used a machine learning approach to identify merger galaxies with HSC-SSP $r$-band morphological parameters.
Their merger fraction is 2\%--3\% in the redshift range of $z\sim$ 0.2--0.7.
Though this is slightly lower than our results (5\%--6\%), our result can be considered to be consistent with theirs given differences of merger definition and the band used.

%Results
\section{Results}
We define the merger fraction $f$ as
\begin{equation}
\label{def_merger_fraction}
f = \frac{N_{\mathrm{merge}}}{N_{\mathrm{tot}}},
\end{equation}
where $N_{\mathrm{tot}}$ and $N_{\mathrm{merge}}$ are the number of inspected galaxies and number of merger galaxies in a $\delta\mathrm{GV}$ bin, respectively.
We count one pair of merging galaxies as a single  ($N_{\mathrm{merge}}=1$).
Uncertainty in  the merger fraction is then estimated through Poisson statistics, $\Delta f=\sqrt{f(1-f)/N_{\mathrm{tot}}}$.

\begin{figure*}
 \begin{center}
  \includegraphics[width=160mm]{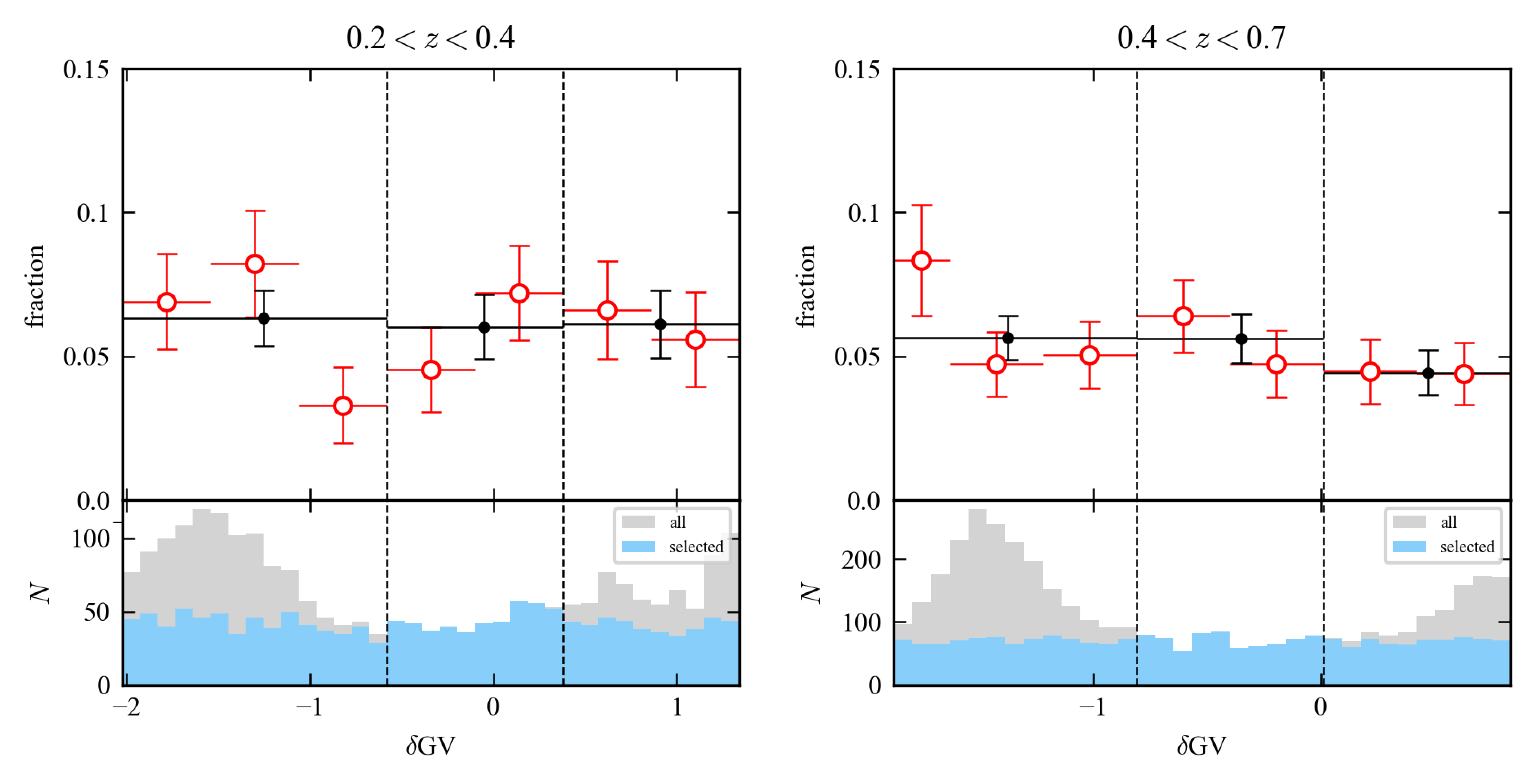}
 \end{center}
  \caption{
Major merger fraction plotted against $\delta$GV.
Left panel shows the results for the redshift range of $z=$ 0.2--0.4 and right panel for $z=$ 0.4--0.7.
The histograms in the lower sub-panels show the number of objects for the whole samples (gray) and selected samples (blue). 
Red open circles show the fractions at a $\delta$GV with uncertainty estimated through Poisson statistics and the bin width indicated by the horizontal red bars.
Black circles show the merger fractions when dividing $\delta\mathrm{GV}$ into 3 bins: SF, GV, and Q galaxy regions.
}
  \label{results_1}
\end{figure*}

Figure \ref{results_1} shows major merger fractions plotted against $\delta$GV (red open circles).  The left panel shows the result for the redshift range of $z=$ 0.2--0.4 and right panel for $z=$ 0.4--0.7.
The histograms in the lower sub-panels show the number of objects for the whole samples (gray) and selected samples (blue). 
The vertical dotted lines indicate the boundaries between SF, GV, and Q galaxies; the area between the lines indicates the GV region, while to the left (right) of the GV region are SF (Q) galaxies.

%figure9

In the redshift range $z=$ 0.2--0.4, the merger fraction is almost constant ($\sim0.06$) with $\delta$GV.
The fraction may appear to be smaller ($\sim 0.04$) than this at $\delta$GV $\sim-0.8$ and $-0.4$, but this is not statistically significant given the uncertainties.
If we plot the major merger fractions for SF, GV, and Q galaxies (black circles in figure \ref{results_1}), the fractions in these three galaxy categories agree with each other within $1\sigma$ uncertainty. In the redshift range $z$=0.4--0.7, the fraction is almost constant around 0.05 over the whole color range.
Although, the merger fraction seems to be larger ($\sim0.08$) than this at the left end of the panel, this excess is again not statistically significant.
As at lower redshifts, differences between the merger fractions in SF, GV, and Q galaxies are less than the 1$\sigma$ uncertainty (black circles).
Therefore, in both redshift ranges, the merger fraction is almost constant around 5\%--6\% without dependence on the galaxy color ($\delta$GV).
This is in contrast to our expectation from the CWD scenario that the major merger fraction should be  higher in the GV region than in SF and Q regions.

\begin{figure*}
 \begin{center}
  \includegraphics[width=160mm]{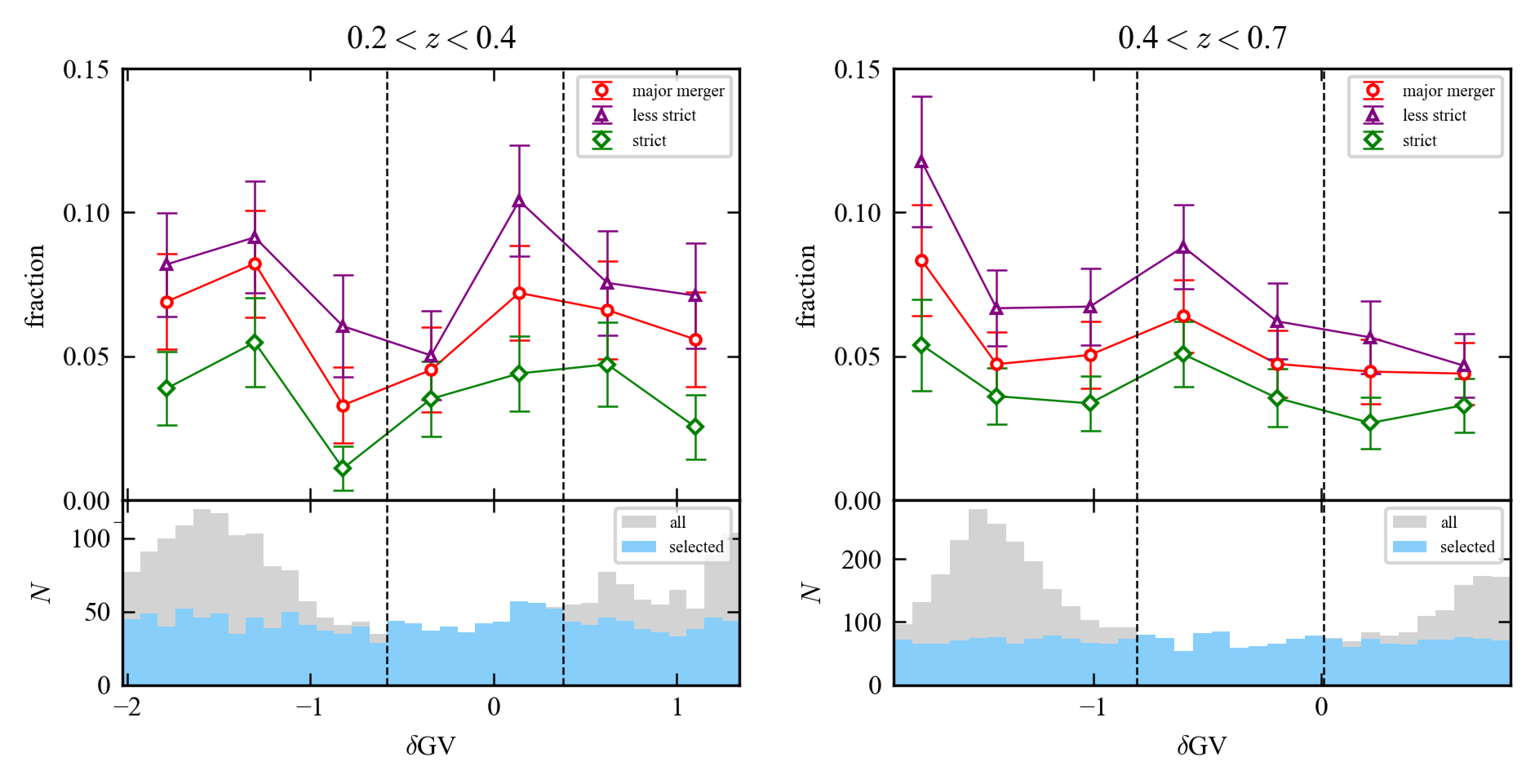}
 \end{center}
\caption{Major merger fractions with the different merger selection criteria.
Left panel shows the results for the redshift range of $z=$ 0.2--0.4 and right panel for $z=$ 0.4--0.7.
Red line and symbols shows the same results as in figure \ref{results_1}.
Purple line and triangles are for the merger fraction with less strict criteria and green line and diamonds are for the fraction with more strict criteria (see text).
}
  \label{frac_discuss}
\end{figure*}

\begin{figure*}
 \begin{center}
  \includegraphics[width=160mm]{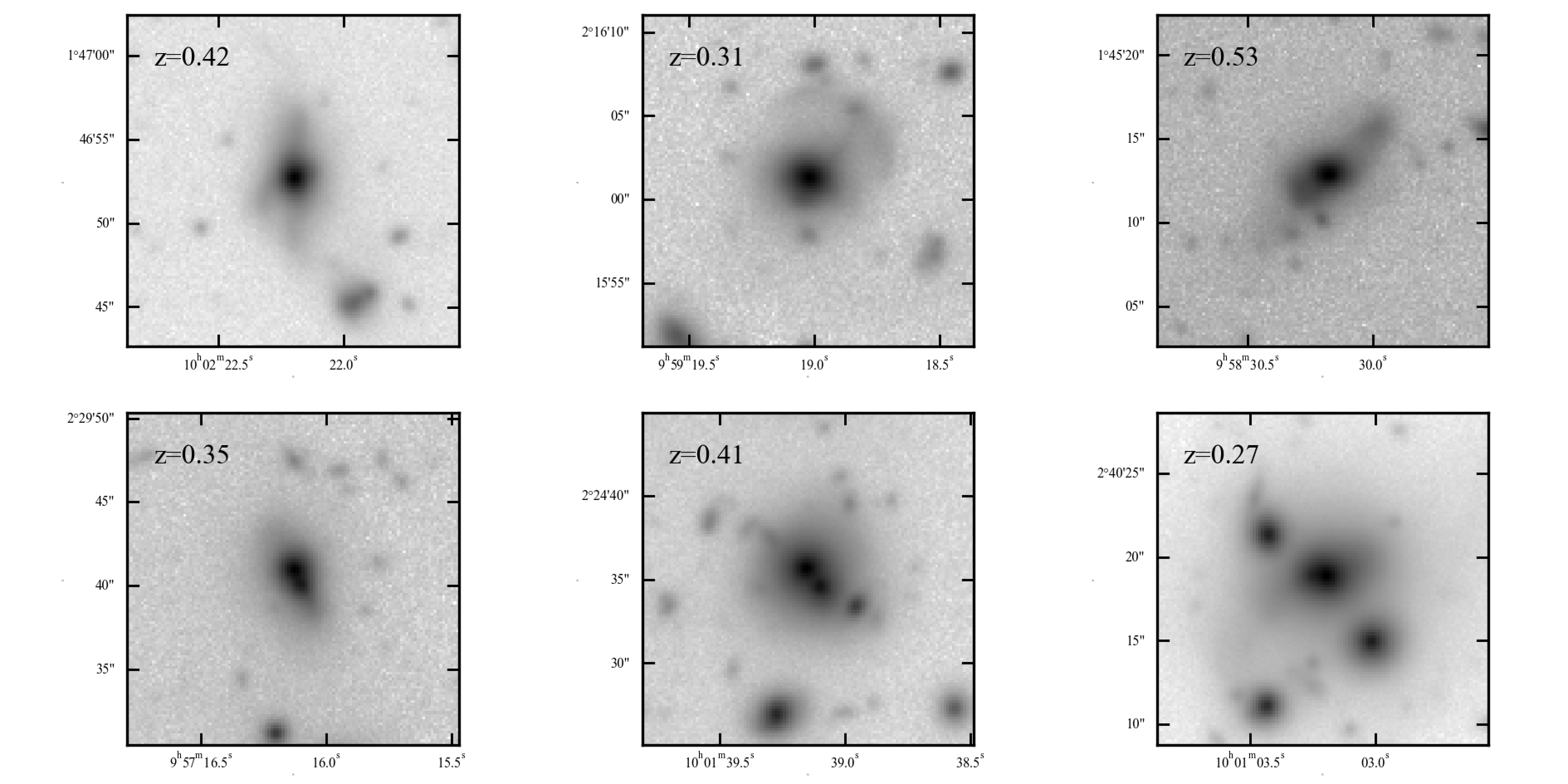}
 \end{center}
 \caption{
Examples of the galaxies included in less strict criteria (upper row) and the galaxies excluded from the samples of major merger in strict criteria (lower row).
North is at the top and east to the left. A field of view of \timeform{20"}$\times$\timeform{20"} is shown.
The photometric redshift of the galaxy is shown at the upper left corner of the each image, and Subaru HSC $i$/$z$-band images are shown if the redshift is in the range of $z=$ 0.2--0.4/$z=$ 0.4--0.7.}
\label{less/strict_cri_major}
\end{figure*}

In order to examine the robustness of the results described above, we examine the major merger fraction against $\delta\mathrm{GV}$ using 1 magnitude brighter sample (i.e., $m_i<21$\>mag for $z=$ 0.2--0.4 and $m_z < 21$\>mag for $z=$ 0.4--0.7).
Resulting distribution of the fractions is very similar to figure \ref{results_1}; no significant change against $\delta$GV is seen.
We further examine the robustness of the results by adopting more/less strict criteria for the identification of major mergers.
The results are shown in Figure \ref{frac_discuss}.
In this figure, red lines and open circles refer to the result shown in figure \ref{results_1}.
Purple lines and triangles show the fractions obtained with less strict criteria;
for these, we amended our original merger sample to also include galaxies that have disturbed structures but have no companion or double nuclei.
These may be galaxies just after a coalescence, but a possibility of minor merger can not be ruled out.
Examples of such galaxies are shown in the upper row of figure \ref{less/strict_cri_major}.
Green lines and diamonds in figure \ref{frac_discuss} show the fractions with more strict merger criteria; here we selected only galaxies with nearly equal angular size companion from our major merger sample.
The lower panels of figure \ref{less/strict_cri_major} shows examples of galaxies that are excluded from the sample of major merger in this way.

Although the absolute values of merger fractions vary with the adopted merger selection criteria, the conclusions remain unchanged.
For the redshift range of $z=$ 0.2--0.4, the overall trend that the merger fraction is almost constant over the whole color range still holds.
This is also the case for the $z=$ 0.4--0.7 redshift range.
Although the fraction for the less strict criteria at $\delta$GV$\sim -1.7$ is slightly elevated (purple triangle in the right panel of figure \ref{frac_discuss}), the difference is only at about $1\sigma$ level and is therefore not significant.
Therefore, our result that the major merger fraction is not significantly higher in the GV region than in the SF and Q regions can be considered to be robust.

%figure10

%figure11

\section{Summary and discussion}
We studied the major merger fraction as function of galaxy color along the massive galaxy quenching channel at $z=$ 0.2--0.7, aiming to examine the Cosmic Web Detachment scenario (CWD; \cite{AragonCalvo+19}). Based on the scenario, the major merger fraction is expected to be high in green valley galaxies as compared with that in star-forming galaxies and quiescent galaxies.
We extracted 1491 (2334) massive ($M_\ast>10^{9.5}$\>$M_\odot$) galaxies with $m_i<22$\>mag ($m_z<22$\>mag) at $z=$ 0.2--0.4 ($z=$ 0.4--0.7) in the rest-frame color range of $0.8<(r-K_s)^0<1.3$ which covers the massive galaxy quenching channel.
Major galaxy merger is defined in our work as a galaxy pair of comparable angular size and luminosity with tidal tails or bridges. Using this criterion, we identified major mergers with visual inspection of deep and good seeing Subaru-HSC-SSP PDR 2 $i$- and $z$-band images to detect faint structures of merging galaxies.
We classified 92 (123) galaxies as major merger galaxies at $z=$ 0.2--0.4 ($z=$ 0.4--0.7).
The resulting major merger fraction is 5\%--6\% and the fraction does not change with the galaxy rest-frame $(\mathrm{NUV}-r)^0$ color.
The fraction is almost the same among blue star-forming galaxies, green valley galaxies, and red quiescent galaxies, even if we change the criteria (relax or tighten) of the major merger.
The result is not consistent with simple expectations based on the Cosmic Web Detachment scenario.

However, our results may not be completely at odds with the CWD scenario given the limitations of our study.
We examine three possibilities here:
that the visibility window during a merger is detectable through morphological features is over by the time the galaxy enters the green valley;
and that CWD quenching imprinting on the merging population is diluted by our lack of ability to trace the cosmic web in this study; 
and that CWD of massive galaxies with $M_\ast>10^{9.5}$ $M_\odot$ is not so frequent in the redshift range we studied.

We discuss these limitations below.

Our study assumed that when a galaxy quenches due to CWD it enters the green valley fairly quickly, while the merger is still in progress.
However, this assumption may be incorrect because the timescales for color evolution may be slower than those we assumed.
For example, \citet{Noirot+22} found that a quenching galaxy can loiter in the SF region of the color-color diagram for quite a long time, $\sim$ 1\>Gyr or more, before it enters the green valley.
Therefore, it is possible that by the time the galaxy enters the green valley, the merger may be over and we will not be able to detect it through the presence of morphological features.
If this is the case then the fraction of mergers in the GV would not be particularly high (meanwhile, CWD-induced mergers, loitering in the SF population, would be diluted in number by the presence of the many regularly star-forming galaxies in that part of color-color space).
Of relevance to this scenario, we note that \citet{Moutard+20} investigated the X-ray hardness ratios of AGNs along the mass-related quenching channel and found that hydrogen column density of AGN in GV corresponds to a typical value for {\it final stage} of gas-poor massive galaxy major mergers.

\begin{figure}
 \begin{center}
  \includegraphics[width=80mm]{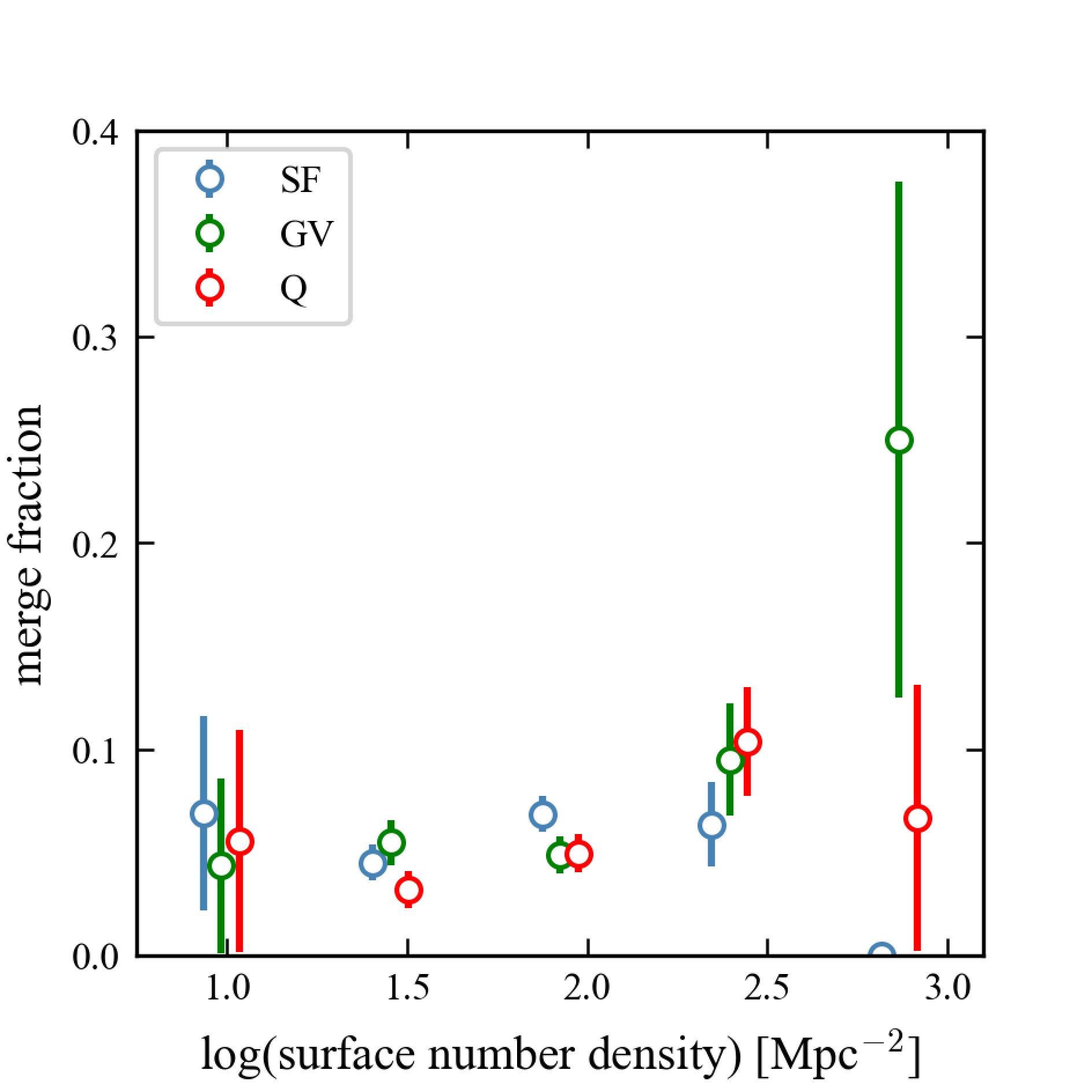}
 \end{center}
 \caption{
Major merger fraction against the surface number density of galaxies. 
Blue, green and red symbols show SF, GV, and Q galaxies respectively, with the uncertainty estimated through poisson statistics. 
 }
 \label{density_merger}
\end{figure}

The other possibility is that merging does happen in the GV as a result of the CWD scenario, but because we are looking at galaxies in all environments (clusters, groups, voids, and in filaments), the signal due to CWD quenching may be diluted.
In order to assess the effect of local density, we examine the merger fraction as function of the local surface number density of galaxies.
We define the surface number density for a galaxy as 
\begin{equation}
\label{surf_num_dense}
\mathrm{surface~number~density} = 6/\pi r_5^2,
\end{equation}
where $r_5$ is the projected physical distance to the 5th closest galaxy.
We calculate $r_5$ using galaxies with $m_i<27$\>mag and in the redshift range $z\pm \sigma_z$ where $\sigma_z = 0.027(1+z)$ and then evaluate the merger fractions in bins of local surface number density.
The resulting merger fractions are shown in figure \ref{density_merger}, where blue, green, and red symbols show the merger fractions of SF, GV, and Q galaxies, respectively.
At the highest surface number density, the merger fraction is higher in GV galaxies than in SF and Q galaxies.
This might indicate that in the denser environment, CWD could actually be happening, though statistical significance is low.
Whether this is caused by CWD quenching or by other effects in very dense regions remains to be seen.
Future work is needed with larger samples to get enough statistics and to trace and characterize cosmic web filaments and other structures to put sample galaxies in their environmental context. 

%figure12

The last possibility is that the web detachment of halo in which a galaxy with $M_\ast > 10^{9.5} M_{\odot}$ resides is not so frequent in the redshift range we studied.
Aragon-Calvo et al. (2019) claim detachment of such halo is the most frequent at $z=$1--2.
Examining the major merger fraction at $z \sim 2$ is desirable, but it is  harder to identify mergers even with the imaging data used in this study.
\citet{AragonCalvo+19} also claim the web detachment events of less massive halo ($M_{\rm halo} \sim 10^{9.5\mathchar`-\mathchar`-10.5}$ $M_\odot$) is much more frequent in the redshift range of 0.2--0.7.
It would be, however, very difficult to identify major merger of dwarf (irregular) galaxy due to their very low surface brightness and to probably their intrinsic irregular morphology.
%In addition, many of such less massive galaxies may reside in a sub-halo in a halo, and it is not so obvious to see whether the merge is due to web detachment or merge in a host halo.  

\bigskip
\begin{ack}
We thank Thibaud Moutrard and the referee for valuable comments which improve the paper. 
K.O. is supported by JSPS KAKENHI Grant Numbers JP19K03928 and JP23K03458.
Y.A. is supported by a Research Fellowship for Young Scientists from JSPS. M.S. thanks the JSPS Invitational Visitor Program for support during a visit to Japan during which we initiated this project.
M.S. also acknowledges support from the Canada Research Chairs program and funding from the NSERC Discovery Grant program and an NSERC Discovery Accelerator.

This paper is based on data collected at the Subaru Telescope and retrieved from the HSC data archive system, which is operated by the Subaru Telescope and Astronomy Data Center (ADC) at National Astronomical Observatory of Japan. Data analysis was in part carried out with the cooperation of Center for Computational Astrophysics (CfCA), National Astronomical Observatory of Japan. The Subaru Telescope is honored and grateful for the opportunity of observing the Universe from Maunakea, which has the cultural, historical and natural significance in Hawaii. 

These data were obtained and processed as part of the CFHT Large Area U-band Deep Survey (CLAUDS), which is a collaboration between astronomers from Canada, France, and China described in \cite{Sawicki+19}.
CLAUDS data products can be accessed from https://www.clauds.net. CLAUDS is based on observations obtained with MegaPrime/ MegaCam, a joint project of CFHT and CEA/DAPNIA, at the CFHT which is operated by the National Research Council (NRC) of Canada, the Institut National des Science de l’Univers of the Centre National de la Recherche Scientifique (CNRS) of France, and the University of Hawaii. CLAUDS uses data obtained in part through the Telescope Access Program (TAP), which has been funded by the National Astronomical Observatories, Chinese Academy of Sciences, and the Special Fund for Astronomy from the Ministry of Finance of China. CLAUDS uses data products from TERAPIX and the Canadian Astronomy Data Centre (CADC) and was carried out using resources from Compute Canada and Canadian Advanced Network For Astrophysical Research (CANFAR).
\end{ack}

%References

\end{document}